\documentclass[hyperref]{article}
\usepackage{graphicx} 
\usepackage{float} 
\usepackage{amssymb}
\usepackage{amsmath}
\usepackage{mathtools}
\usepackage[thmmarks,amsmath]{ntheorem} 
\usepackage{bm} 
\usepackage{extarrows}
\usepackage[a4paper]{geometry}                           
\usepackage{bibspacing}
\date{}
\usepackage[marginal]{footmisc}

\usepackage{graphicx}
\usepackage{subfigure}

\begin{document}

\title{\textbf{MutualGraphNet: A novel model for motor imagery classification}}
\author{\sffamily Yan Li$^1$, Ning Zhong$^{2}$, David Taniar${^3}$, Haolan Zhang$^{4,}$\footnote{corresponding author: haolan.zhang@{nit.zju.edu.cn}}\\
{\sffamily\small $^1$ Zhejiang University, China, }
{\sffamily\small $^2$ Maebashi Institute of Technology, Japan}\\
{\sffamily\small $^3$ Monash University, Australia, }
{\sffamily\small $^4$ Ningbo Research Institute, Zhejiang University, China}\\
{\sffamily\small ly21121@zju.edu.cn, zhong@maebashi-it.ac.jp, David.Taniar@monash.edu, haolan.zhang@nit.zju.edu.cn}
}

\maketitle

\noindent{\bf Abstract: }
Motor imagery classification is of great significance to humans with mobility impairments, and how to extract and utilize the effective features from motor imagery electroencephalogram(EEG) channels has always been the focus of attention. There are many different methods for the motor imagery classification, but the limited understanding on human brain requires more effective methods for extracting the features of EEG data. Graph neural networks(GNNs) have demonstrated its effectiveness in classifying graph structures; and the use of GNN provides new possibilities for brain structure connection feature extraction. In this paper we propose a novel graph neural network based on the mutual information of the raw EEG channels called MutualGraphNet. We use the mutual information as the adjacency matrix combined with the spatial temporal graph convolution network(ST-GCN) could extract the transition rules of the motor imagery electroencephalogram(EEG) channels data more effectively. Experiments are conducted on motor imagery EEG data set and we compare our model with the current state-of-the-art approaches and the results suggest that MutualGraphNet is robust enough to learn the interpretable features and outperforms the current state-of-the-art methods.



\section{Introduction}

Brain-computer-interface(BCI) technology has received much attention globally because of its significant meaning\cite{Song19}. It enables their users to interact with a machine via the brain singals\cite{Wolpaw02}, such as the motor imagery task which translates the mental imagination of movement into commands\cite{Blankertz07} can be utilized to help disabled people as a rehabilitation device\cite{Wolpaw91}, which could be considered as being the only way of communication for people with motor disabilities\cite{Brain01}. The motor imagery classification based on the features extracted from the EEG imagination data of moving the body parts without actual movement, but the feature extraction process often rely too much on prior knowledge to exclude some features\cite{Mcfarland06}, so the more robust feature extraction techniques will continue to promote the development of BCI technologies.

A complete BCI system generally contains four main processes\cite{Lotte2018}: EEG raw data collection, data pre-processing, feature extraction and feature classification. 
The classification phase is crucial since a good classifier can utilize the extracted features as great as possible and greatly improve the accuracy of classification. The motor imagery classification is a EEG-based task mainly focus on the feature extraction and classification, and a big number of researchers have made great contributions to this task. The two most common types of features are frequency band power features and time point features\cite{Makeig12}, both types of features benefit from being extracted after spatial filtering\cite{Lotte14}. Principal component analysis(PCA) and independent component analysis(ICA) are two classical unsupervised spatial filter manners\cite{Kachenoura08}, supervised spatial filters include the common spatial patterns(CSP) and filter bank common spatial patterns(FBCSP)\cite{Keng12}. As for the classifiers for Motor imagery task, there are some state-of-the-art methods that have been proven to have good performance, such as linear discriminant analysis(LDA), passive-aggressive(PA) and support vector machine(SVM)\cite{Woehrle15}.

In recent years, deep learning methods have achieved remarkable performances in many fields, and many recent works have explored the application of deep learning to EEG-based tasks\cite{Schirrmeister17}. Deep learning has largely alleviated the need for manual feature extraction and provided end-to-end learning for EEG based tasks, such as sleep stage detection, anomaly detection, motor imagery classification and so on\cite{Lawhern16}. Although the convolution nets could learn from the raw data without manual feature extraction, but they still have their own limitations including that they generally require large dataset to train the model, which may be a drawback for EEG based tasks since that the collection of EEG data often cost a lot. In addition, the EEG data has its own unique characteristics, it's collected from different areas of the brain, so the spatial connection among the EEG data shouldn't be ignored, however, neither the state-of-the-art methods nor the recent deep learning methods can effectively learn the connections between different brain regions\cite{Jia20}.

Graphs are the most appropriate data structure to indicate brain connection, and graph neural networks(GNNs) has been demonstrated to be effective in classifying graph structures\cite{Zhou20}, the core idea of GNNs is to update each node's embedding iteratively through aggregating the representations of its neighbors and itself. The EEG channels could be represented as nodes in the graph and the connections among the channels correspond to the edges of the graph, but the graph convolutional networks need adjacency matrix to be given in advance which is the representation of the graph connection\cite{Bruna13}, since the limitations of cognition of brain structure, so it's still a challenge to determine a suitable graph structure of the brain. Moreover, the collection of EEG data is usually in chronological order, so it also has temporal characteristics in addition to spatial characteristics which also need to be taken into account.

In this paper, we proposed a novel model called MutualGraphNet, combined the spatial-temporal filter and graph convolutional networks to learn the temporal and spatial characteristics, as for the graph connection, we use the mutual inforamtion between different channels as the degree of the association between channels, which achieved robust performance on the motor imagery classification tasks. The contributions of this paper are as follows:
\begin{itemize}
  \item A novel model is proposed, which could realize end-to-end learning. Furthermore, the model is specially designed to adapt to the characteristics of EEG data, so it could be able to utilize the features to the great extend.
  \item For the first time, we try to use mutual information as the adjacent matrix of graph convolution input, and achieve good performance.
  \item Experimental results demonstrate that our method achieves better performance in the motor imagery classification task.
\end{itemize}

\section{Related Work}

A motor imagery classification task is of great significance for disabled people. Numerous work has been proposed to improve the  classification performance. In the earlier research, traditional machine learning methods are generally used for motor imagery classification task, such as support vector machine(SVM), K-Nearest-Neighbor(KNN) and artificial neural network(ANN) are frequently used\cite{Halta19}, but these traditional methods have limited performance on EEG-based classification tasks. Thus, Common Spatial Patterns(CSP)\cite{J99Designing} algorithm was proposed specially for motor imagery classification tasks, and it has been used for many EEG-based classification tasks successfully. Based on the CSP, the Filter bank common spatial pattern(FBCSP)\cite{Kai08} was proposed, which bandpass-filtered the EEG data into multiple frequency bands, then CSP features were extracted from each of these bands. The FBCSP algorithm achieved better performance than CSP on the motor imagery classification task.

Currently, the deep learning methods are utilized in EEG-based classification tasks, Deep Belief Network(DBN)\cite{Lu16} was proposed that extracted features manually from the channels then feed the features into the network. Convolutional Neural Networks(CNN) could learn features from EEG data automatically and have better performance than DBN due to its regularization structure and degree of translation invariance\cite{Aggarwal19}. Two CNN models were specially designed for motor imagery classification called Shallow ConvNets and Deep ConvNets\cite{Schirrmeister17}, both of them have better performance than the state-of-the-art methods. Then another CNN model called EEGNet\cite{Lawhern16} was proposed, which utilizes the Depthwise and Separable convolutions to replace the traditional convolutions for the motor imagery task. Besides, the novel deep learning method based on Multi-Task Learning framework called DMTL-BCI\cite{Song19} was proposed for EEG-based classification, which enhanced the generalization ability of the model and improve the performance of classification with limited data.

The CNN models can effectively extract the local patterns of data, but it can only be applied for the standard grid data\cite{Guo19}, graph convolutional networks have been proven to have better performance on the graph structure data. Many works have been done to improve the performance of the graph convolutional networks. For example, The GraphSAGE model\cite{Hamilton17} could leverage node feature information to efficiently generate node embeddings for unseen data, the graph attention networks(GATs)\cite{Velikovi17} through stacking layers in which nodes are able to attend over their neighborhoods' features, could specify different weights to different nodes in a neighborhood. So far, GCNs have been applied in many fields, the spatial-temporal graph convolution network(ST-GCN)\cite{Li18} is proposed to learn the dynamic graphs for the human action recognition tasks, the spatiotemporal multi-graph convolution network(ST-MGCN)\cite{Geng19} is proposed for ride-hailing demand forecasting which encodes the non-Euclidean correlations among regions into multiple graphs, GraphSleepNet\cite{Jia20} based on spatial-temporal convolution network(ST-GCN) is proposed for automatic sleep stage classification.

Motivated by the studies mentioned above, considering the graph structure and the dynamic
spatial-temporal characteristic of the EEG data, also the graph structure of different motor imagery may be different, the traditional GCNs models may be not optimal for EEG-based motor imagery classification task. we propose the novel model to best suit to the characteristics of EEG data.
%

\section{Methodology}
The overall framework of the model proposed in this paper is presented in Fig. 1, it includes three main parts: feature extraction and mutual information computation part, spatial-temporal attention part and spatial-temporal graph convolution part. Spatial-temporal attention part puts more attention on the more valuable spatial-temporal information, spatial-temporal graph convolution part extracts both spatial and temporal features.
\begin{figure}[ht]
\centering
\includegraphics[scale=0.6]{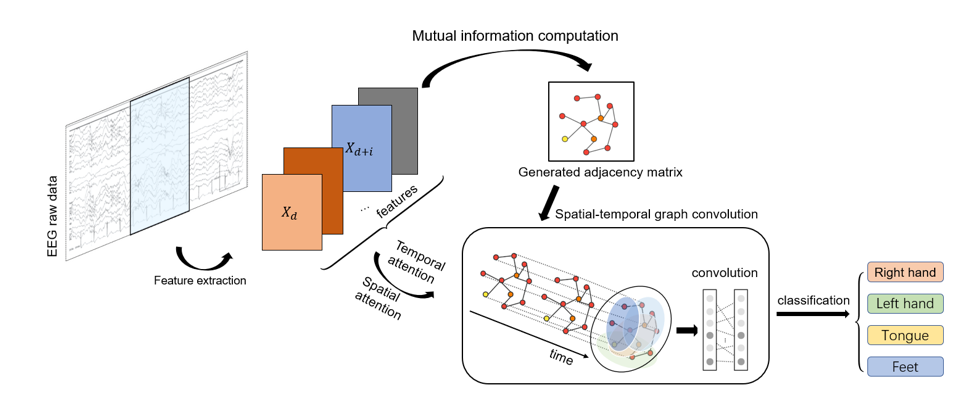}
\caption{The overall structure of the proposed model, consists of three parts : the feature extraction and mutual information computation part, the spatial-temporal attention mechanism part and spatial-temporal graph convolution part.}
\label{fig:label}
\end{figure}
\subsection{Mutual information computation}
Mutual Information(MI)\cite{1997Multimodality} is used to indicate wether there is a relationship between two variables and the strength of the relationship. The mutual information of two variables $X$ and $Y$ can be defined as:
\begin{equation}
  I(X,Y) = \sum_{x\in X}\sum_{y\in Y}p(x,y)log\frac{p(x,y)}{p(x)p(y)}
\end{equation}
Mutual information is related to entropy, which is the expected or mean value of the information of all variables. The entropy of X is defined as:
\begin{equation}
  H(X) = \sum_{x\in X}P(x)log\frac{1}{P(x)} = - \sum_{x\in X}P(x)logP(x) = -ElogP(X)
\end{equation}
 Then MI of $X$ and $Y$ can be computed by the equations:
\begin{equation}
   I(X,Y) = H(X) + H(Y) -H(X,Y) = H(X)-H(X|Y) = H(Y)-H(Y|X)
\end{equation}
$H(X,Y)$ is the joint entropy of $X$ and $Y$:
\begin{equation}
H(X,Y) = \sum_{x \in X}\sum_{y \in Y}p(x,y)log\frac{1}{p(x,y)} = -ElogP(X,Y)
\end{equation}
$H(Y|X)$ is the conditional entropy that $X$ is given in advanced:
\begin{equation}
H(Y|X) = \sum_{x \in X}\sum_{y \in Y}p(x)p(y|x)log\frac{1}{p(y|x)} = -ElogP(Y|X)
\end{equation}
Thus, $I(X,Y)$ is the reduction in the uncertainty of the variable X by the knowledge of another variable Y, equivalently, it represents the amount of information that B contains about A.

Considering the features of EEG data $X = {x^1, x^2,...,x^N}\in\mathbb{R}^{N\times F}$, we could compute the mutual information $m_ij$ of $x^i$, $x^j$ and use it as the weight of the connection of $x^i,x^j$, then we could generate a $N\times N$ weight matrix which could be used as the input adjacency matrix of the graph convolution networks as Fig. 2 shown.
\begin{figure}[ht]
\centering
\includegraphics[scale=0.4]{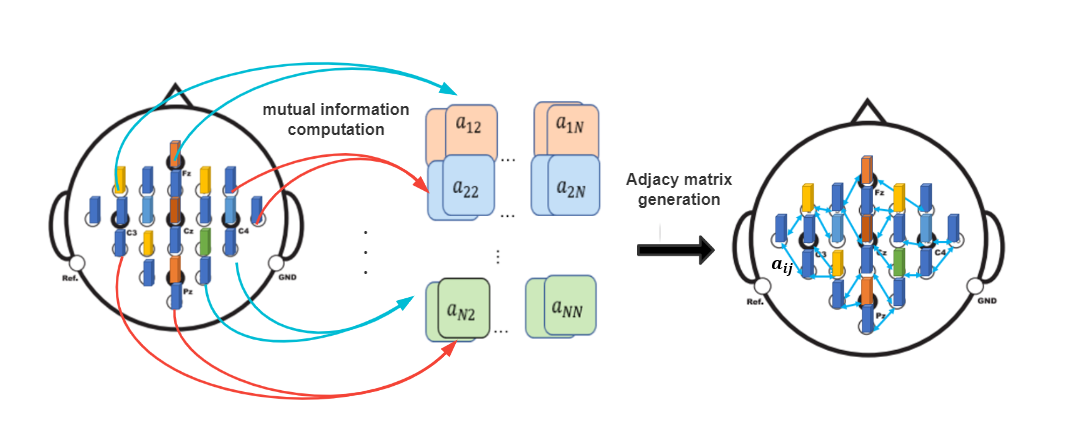}
\caption{The process of calculating the mutual information between channels and obtaining the adjacency matrix, $N$ represents the number of channels.}
\label{fig:label}
\end{figure}

\subsection{Spatial-temporal attention}
The spatial-temporal attention mechanism could capture the dynamic spatial and temporal correlations of the motor imagery network. In the spatial dimension, the activities of one brain region has influence on other brain regions and generally different brain activities convey different information, so the dynamic spatiotemporal capture mechanism is required. We use a spatial  attention mechanism\cite{2017Effective}, which could be represented as:
\begin{equation}
S = V_p*\sigma((\chi^{(r-1)}W_1)W_2(W_3\chi^{(r-1)})^T +b_{p})
\end{equation}
\begin{equation}
S^{'}_{i,j} = \frac{exp(S_{i,j})}{\sum^N_{j=1}exp(S_{i,j})}
\end{equation}

Where $S$ denotes the spatial attention matrix, which is computed by current layer. $ V_p, b_p \in \mathbb{R}^{N\times N}$, $\chi^(r-1) = (X_1,X_2, ... , X_{T_{r-1}} \in \mathbb{R}^{N\times C_{r-1}\times T_{r-1}}$ $C_{r-1}$ is the number of channels of the input data in the $r^th$ layer. $W_1 \in \mathbb{R}^{T_{r-1}}, W_2 \in \mathbb{R}^{C_{r-1}\times T_{r-1}}, W_3 \in \mathbb{R}^{C_{r-1}}$, $S_{i,j}$ in $S$ represents the correlation strength between node $i$ and $j$, then a softmax function is used to normalize the attention weights. Combine the adjacency matrix and the spatial attention matrix, the model could adjust the impacting weights between nodes dynamically.

In the temporal dimension, there are correlations during each motor imagery trial, since that the brain waves are transmitted in the cerebral cortex and the active areas of the brain will change over time, so the collected EEG data also changes over time. Therefore, a temporal attention is utilized to capture dynamic temporal information. The temporal attention mechanism is defined as:
\begin{equation}
E = V_e*\sigma(((\chi^{(l-1)})^{T}M_1)M_2(M_3\chi^{(l-1)})+b_q)
\end{equation}
\begin{equation}
E^{'}_{m,n} = \frac{exp(E_{i,j})}{\sum^{T_{r-1}}_{j=1}exp(E_{i,j})}
\end{equation}
Where $V_e, b_q \in \mathbb{R}^{T_{l-1}\times T_{l-1}},  M_1 \in \mathbb{R}^N, M_2 \in \mathbb{R}^{C_{l-1}\times N}, M_3 \in \mathbb{R}^{C_{l-1}}$, $E_{m,n}$ denotes the strength of the correlation between motor imagery network $m,n$, and $E$ is normalized by the softmax function, so the temporal attention matrix can be directly applied to the input.
\subsection{Spatial-temporal graph convolution}

The spatial-temporal convolution consists of a graph convolution in the spatial dimension and a normal convolution in the temporal dimension, which could extract both the spatial features and the temporal features.

The spatial features are extracted by aggregating information from neighbor nodes, we use graph convolution to extract the spatial features. The graph convolution is based on laplacian matrix and fourier transform, the graph laplacian can be defined as:
\begin{equation}
  L = I - D^{-1/2}AD^{-1/2}
\end{equation}
Where $A\in \mathbb{R}^{N\times N}$ is the adjacency matrix associated with the graph, $D \in \mathbb{R}^{N\times N}$ is the diagonal degree matrix, $I\in \mathbb{R}^{N\times N}$ is the identity matrix. $L$ is a real symmetric positive semidefinite matrix, it can be decomposed as $ L = U\Lambda U^T$ and $\Lambda \in \mathbb{R}^{N\times N}$ is the diagonal matrix of eigenvalues that represent the frequencies of their associated eigenvectors. Let $x \in \mathbb{R}^n$ be a signal defined on the vertices of a graph $G$, the graph fourier transform of the signal is defined as \^{x} = $U^Tx$. The graph convolution uses the linear operators that diagonalize in the flourier domain to replace the classical convolution operator, the graph convolution can be defined as:
\begin{equation}
  g_\theta(L)x = g_\theta(U\Lambda U^T)x = Ug_\theta(\Lambda)U^Tx
\end{equation}
Where $\theta$ is a vector of fourier coefficients to be learned, $g_\theta$ is the filter, to reduce the computational complexity, $g_\theta$ can be approximated by a truncated expansion in the terms of Chebyshev polynomials\cite{2016Convolutional}:
\begin{equation}
 g_{\theta}({\Lambda}) = \sum^{k-1}_{p=0}{\theta}_{p}T_{p}(\tilde{\Lambda})
\end{equation}
Where $k$ is the order of the Chebyshev polynomials, $\theta_p \in \mathbb{R}^k$ is the vector of Chebyshev coefficients, $T_{p}(\tilde{\Lambda})\in \mathbb{R}^{N\times N}$ is the Chebyshev polynomial of order $k$ and $ \tilde{\Lambda} = 2{\Lambda}/{\lambda}_{max} -I$ ranges in $[-1,1]$. Then the $j-th$ output feature can be calculated as:
\begin{equation}
y_i = \sum^{F_{in}}_{i=1}g\theta_{i,j}(L)x_i
\end{equation}
 $x_i$ denotes the $i-th$ row of input matrix, $F_{in}$ equals to the input dimension, the outputs are collected into a feature matrix $Y = [y_1,y_2,...,y_{F_{out}}] \in \mathbb{R}^{N \times F_{out}}$. In this work, we generalize the above definition to the nodes with multiple channels, the $l-th$ layer's input is $X^{(l-1)} = (x_1,x_2,...,x_{(T_{l-1})}) \in \mathbb{R}^{N\times C_{l-1}\times T_{l-1}}$, $C_{(l-1)}$ denotes the channel's number and $T_{l-1}$ denotes the $l-th$ layer's temporal dimension.

After the graph convolution having captured the neighboring information for each node in the spatial dimension, a standard convolution layer is used in the temporal dimension, we use a standard two-dimension convolution layer to extract the temporal information, the $r-th$ convolution layer could be defined as:
\begin{equation}
\chi^{(r)}_h = ReLU(\Phi*(ReLU(g_\theta * G \hat{\chi}^{(r-1)}_{h})))
\end{equation}
where $\Phi$ is the parameter of the temporal dimension convolution kernel, and $*$ represents the convolution operation, ReLU is the activation function.
\section{Experiment}
In order to evaluate the effectiveness of our model, we carried out the comparative experiments on a public dataset BCI Competition IV dataset 2a(SMR) for motor imagery task.
\subsection{Dataset description}
The BCI Competition IV dataset 2a consists of EEG data from nine subjects, there are two sessions recorded, one for training and the other one for testing. Each session includes 288 trials, which are recorded with 22 EEG electrodes and 3 electrooculogram channels, and we only utilize the 22 EEG channels in this experiment. There are four type of labels in this dataset, corresponding to movements of the left hand, right hand, feet and tongue.

The original dataset is sampled at 250Hz and bandpass-filtered between 0.5Hz and 100Hz, and we low-pass filter the dataset to 4-40Hz. Also in our experiment, we set the length of each trial to 4.5s which starts from 500ms before the start cue of each trial until to the end cue, then extract 11 differential entropy features(DE) for each channel then double fold the features to make it have the same shape as the adjacency matrix, and combine the two as the input of the graph convolutional network, then we standard scale the data to make it suit for the machine learning model. To show the effectiveness of our model learning from the raw data and ensure the model could be used for wider range of task , we don't do much more preprocessing on the raw EEG data.
\subsection{Experiment settings}
To show the effectiveness of our model, we compare our model with some state-of-the-art methods, the baseline methods are listed as follows:
\begin{enumerate}
  \item Filter Bank Common Spatial Patterns(FBCSP)\cite{2012Filter}: It extracted the band power features of EEG ,then use the features to train the classifier to predict the labels.
  \item Shallow ConvNet\cite{Schirrmeister17}: An end-to-end learn method, which use convolutional networks to do all the computations.
  \item Deep ConvNet\cite{Schirrmeister17}: It has more convolution-pooling blocks and is much deeper than Shallow ConvNet.
  \item EEGNet\cite{Lawhern16}: It uses the depthwise and separable convolution and has two convolution-pooling blocks.
\end{enumerate}
In addition to the above baseline methods, we also compared traditional machine learning methods, support vector machine(SVM)\cite{2007LIBSVM} and random forest(RF)\cite{2002Classification}.

In order to prove that the model can effectively extract features and have the ability to eliminate the influence of individual differences, we no longer conduct experiments on each subject separately, we mixed the experimental data of nine subjects, and a total 2592 training trials and 2592 testing trials, and we use four-fold cross-validation to evaluate the performance. We use Adam optimizer and the loss rate is set to 7.6e-4, and dropout rate is 0.5, the batch size is 32 and the we train the model 500 epochs. For the graph convolutional layers, and since the training set is not big enough, so in order to reduce the impact of overfitting, we adopt a loss flooding strategy\cite{2020Do}, which is defined as:$ \tilde{R}(g) = |\tilde{R}(g) - b| + b $ and $\tilde{R}(g)$ is the loss of the model, $b$ is a constant called loss flooding level, here we set $b$ as 0.5. 
As for the baseline methods, in order to evaluate the performance of the models more reasonably, we use 250Hz sampling 4.5s EEG data for all experiments. Since that the EEGNet\cite{Lawhern16} used the 128Hz resampled data to conduct experiment in the original paper, so we double the lengths of temporal kernels and average pooling size of the original model for double sampling rate to better adapt the input which proven to have better performance than the original model. In response to changes in the length of the sampling time, we also adjusted the parameters of each model accordingly and conducted experiments and we selected the best model performance. The training parameter of other baseline methods are the same as in the paper \cite{Lawhern16}.

\subsection{Results and discussion}
We compare our model with the six baseline methods on SMR, we use the accuracy, f1-score and precision as the evaluation metrics to evaluate the performance of the models. Tabel 1 shows the performance of the different models on the SMR dataset, the results show that our model have better performance compared to the other baseline methods.
\begin{table}[ht]
\centering
\caption{The performance comparison of the state-of-the-art approaches on the SMR dataset}
\begin{tabular}{lccr}
  \hline
  \hline
  Model&Accuracy&F1-score&Precision\\
  \hline
  SVM &0.3488 &0.3485 &0.3486 \\
  Deep ConvNet & 0.3507 & 0.3191 & 0.4148 \\
  FBCSP & 0.3511 & 0.3366 & 0.3714 \\
  RF &0.4008 &0.3996 &0.4004 \\
  EEGNet &0.4616  & 0.4838 & 0.5095 \\
  Shallow ConvNet & 0.4857 & 0.4789 & 0.4978 \\
  MutualGraphNet &\textbf{0.5190} &\textbf{0.5175} &\textbf{0.5208} \\
  \hline
  \hline
\end{tabular}
\end{table}
For the traditional methods, the random tree(RF) has better performance than the support vector machine(SVM), but both of them aren't good enough. The FBCSP cannot extract and utilize complex features in multi-subject tasks, though it has good performance in single-subject tasks. And the results show that the traditional machine learning methods can't learn the complex features well, the deep learning models EEGNet and ShallowConvNet all outperform the traditional methods which demonstrate the effectiveness of deep convolutional neural networks for EEG-bask classification tasks, however the performance of DeepConvNet demonstrates that the deeper convolutional network doesn't work better.

In order to evaluate the effect of the depth of network, we study the impact of the layers of ST-GCN in Fig. 3. The horizonal axis in Fig. 3 represents the layers of ST-GCN and the vertical axis represents the corresponding performance of the model. The results show that the MutualGraphNet with more ST-GCN layers doesn't work better, the best performance is achieved when the layers is 4 and with the number of layers increases the performance gets worse, which is caused by the increase in the number of layers leads to an increase in training parameters, but the training data set is too small to train the model with more parameters.
\begin{figure}[h]
    \centering
    \begin{minipage}[t]{0.6\linewidth}
		\centering
		\includegraphics[scale=0.6]{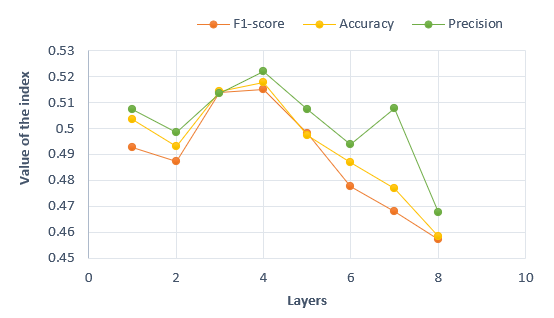}
		\caption{Performance of model with different ST-GCN layers.}
        \label{fig:label}
	\end{minipage}
\end{figure}
In this paper, we extract differential entropy(DE) feature as the input of the model, and in EEG-based tasks there are other five different features\cite{2016Identifying}: power spectral density(PSD), differential asymmetry (DASM), rational asymmetry(RASM), asymmetry(ASM) and differential caudality(DACU) features from EEG. The DASM and RASM can be expressed as:
\begin{equation}
 DASM = DE(X_{left}) - DE(X_{right})
\end{equation}
\begin{equation}
 RASM = DE(X_{left})/ DE(X_{right})
\end{equation}
ASM features are the direct concatenation of DASM and RASM features. DCAU features are the difference between DE features of frontal-posterior electrodes, which can be defined as:
\begin{equation}
 DCAU = DE(X_{frontal})-DE(X_{posterior})
\end{equation}
 We also evaluate the performance of our model on these features. The comparison of the performance of model on different features is shown in Table.2. All the experiments are performed with 4-fold cross-validation and the training settings are the same as above.
\begin{table}[ht]
\centering
\caption{The performance of model for different features}
 \begin{tabular}{lccr}
 \hline
 \hline
  Feature& Accuracy& F1-score& Precision\\
  \hline
  PSD& 0.2604&0.2286& 0.2595\\
  DSAM&0.3646&0.3523&0.3541\\
  ASM&0.3815&0.3820&0.3879\\
  ASDM&0.3811&0.3777&0.3764\\
  DCAU&0.4162&0.4144&0.4191\\
  DE &\textbf{0.5190} &\textbf{0.5175} &\textbf{0.5208}\\
  \hline
  \hline
 \end{tabular}
\end{table}
The results presented in Table.2 show that DE features outperform the other features and DCAU features also achieve comparable performance, but ASDM and DSAM features contains less information which leads to limited performance. The results indicate that there exists some kind asymmetry of the brain which has discriminative information and our knowledge of the human brain is still very limited, the deeper understanding of brain is still required to obtain more effective and valuable information from EEG data. Moreover, in order to evaluate the effectiveness of mutual information(MI), we also design several different ways to generate adjacency matrix:
\begin{enumerate}
  \item KNN: For each channel, select the nearest $N$ channels to establish a connection.
  \item Euclidean distance(ED): According to the actual distance of each electrode on the brain, select adjacent points to establish a connection.
  \item Random: Randomly select channels and establish connections between channels.
  \item Mul\_KNN: Use KNN to establish connections and calculate mutual information between connected channels.
  \item Mul\_ED: Use euclidean distance to conform connection and calculate mutual information between connected channels
\end{enumerate}
The results of classification with different kind of adjacency matrix are shown in Fig. 4.
\begin{figure}[h]
\centering
\begin{minipage}[t]{0.6\linewidth}
		\centering
		\includegraphics[scale=0.6]{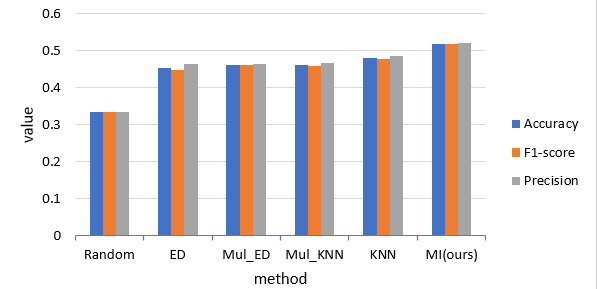}
		\caption{The performance of model for different features.}
        \label{fig:label}
	\end{minipage}
\end{figure}
It can be seen that the MI adjacency matrix has best performance, Mul\_KNN and Mul\_ED is better than KNN and ED which means that mutual information could provide valuable information for ST-GCN. Furthermore, the adjacency matrix surly could effect the performance of classification.

\section{Conclusion}
In this paper, we propose a novel attention based spatial-temporal graph convolution model for motor imagery classification task, the model combines the spatial-temporal attention mechanism and the spatial-temporal convolution. The main advantage of the proposed model is to extract attentive spatial-temporal features of EEG data and experiment results demonstrate that our model outperforms the state-of-the-art methods. Our model is designed to extract spatial-temporal features of EEG, so it can also be applied to other EEG-based tasks.

Actually, the current understanding of brain mechanisms is still limited, as the understanding of the brain gradually increase, more influencing factors will be taken into account to further improve the forecasting accuracy. Moreover, motor imagery EEG data presents individual differences, such as FBCSP has different performance when experimenting with EEG data from different subjects, and it can achieve good results when using the same subject's data for training and testing, but it does not perform well in mixed data of multiple subjects. Individual difference also affected the development of solutions for the classification task of motor imagery, eliminating individual differences and extract valuable features is still key for wider application of EEG-based tasks.

\section{Ackowledgment}
This work is partially supported by Humanity and Social Science Foundation of the Ministry of Education of China (21A13022003), Zhejiang Provincial Natural Science Fund (LY19F030010), Zhejiang Provincial Social Science Fund (20NDJC216YB), Ningbo Natural Science Fund (No. 2019A610083), Zhejiang Provincial Educational Science Scheme 2021 (GH2021642) and National Natural Science Foundation of China Grant (No.72071049).
\bibliographystyle{plain}
\bibliography{reference}

\begin{thebibliography}{10}

\bibitem{Aggarwal19}
S.~Aggarwal and N.~Chugh.
\newblock Signal processing techniques for motor imagery brain computer
  interface: A review.
\newblock {\em Array}, s 1–2.

\bibitem{Blankertz07}
B.~Blankertz, G.~Dornhege, M.~Krauledat, KR~Müller, and G.~Curio.
\newblock The non-invasive berlin brain-computer interface: Fast acquisition of
  effective performance in untrained subjects.
\newblock {\em NeuroImage}, 37(2):539--550, 2007.

\bibitem{Bruna13}
J.~Bruna, W.~Zaremba, A.~Szlam, and Y.~Lecun.
\newblock Spectral networks and locally connected networks on graphs.
\newblock {\em Computer Science}, 2013.

\bibitem{2007LIBSVM}
C.~C. Chang and C.~J. Lin.
\newblock Libsvm: A library for support vector machines.
\newblock {\em ACM Transactions on Intelligent Systems and Technology}, 2(3,
  article 27), 2007.

\bibitem{2016Convolutional}
Michal Defferrard, X.~Bresson, and P.~Vandergheynst.
\newblock Convolutional neural networks on graphs with fast localized spectral
  filtering.
\newblock 2016.

\bibitem{Lotte14}
Lotte F.
\newblock A tutorial on eeg signal processing techniques for mental state
  recognition in brain-computer interfaces.
\newblock {\em Springer London}, 2014.

\bibitem{2017Effective}
X.~Feng, G.~Jiang, Q.~Bing, T.~Liu, and Y.~Liu.
\newblock Effective deep memory networks for distant supervised relation
  extraction.
\newblock In {\em Twenty-Sixth International Joint Conference on Artificial
  Intelligence}, 2017.

\bibitem{Geng19}
X.~Geng, Y.~Li, L.~Wang, L.~Zhang, and Y.~Liu.
\newblock Spatiotemporal multi-graph convolution network for ride-hailing
  demand forecasting.
\newblock {\em Proceedings of the AAAI Conference on Artificial Intelligence},
  33:3656--3663, 2019.

\bibitem{Guo19}
Shengnan Guo, Youfang Lin, Ning Feng, Chao Song, and Huaiyu Wan.
\newblock Attention based spatial-temporal graph convolutional networks for
  traffic flow forecasting.
\newblock {\em Proceedings of the AAAI Conference on Artificial Intelligence},
  33:922--929, 07 2019.

\bibitem{Halta19}
Kadir Haltaş, Atilla Erguzen, and Erdal Erdal.
\newblock Classification methods in eeg based motor imagery bci systems.
\newblock pages 1--5, 10 2019.

\bibitem{Hamilton17}
William Hamilton, Rex Ying, and Jure Leskovec.
\newblock Inductive representation learning on large graphs.
\newblock 06 2017.

\bibitem{2020Do}
T.~Ishida, I.~Yamane, T.~Sakai, G.~Niu, and M.~Sugiyama.
\newblock Do we need zero training loss after achieving zero training error?
\newblock 2020.

\bibitem{Jia20}
Z.~Jia, Y.~Lin, J.~Wang, R.~Zhou, and Y.~Zhao.
\newblock Graphsleepnet: Adaptive spatial-temporal graph convolutional networks
  for sleep stage classification.
\newblock In {\em Twenty-Ninth International Joint Conference on Artificial
  Intelligence and Seventeenth Pacific Rim International Conference on
  Artificial Intelligence {IJCAI-PRICAI-20}}, 2020.

\bibitem{Kachenoura08}
A.~Kachenoura, L.~Albera, L.~Senhadji, and P.~Comon.
\newblock Ica: A potential tool for bci systems.
\newblock {\em Signal Processing Magazine IEEE}, 25(1):57--68, 2008.

\bibitem{Kai08}
K.~A. Kai, Y.~C. Zhang, H.~Zhang, and C.~Guan.
\newblock Filter bank common spatial pattern (fbcsp) in brain-computer
  interface.
\newblock In {\em IEEE International Joint Conference on Neural Networks},
  2008.

\bibitem{Brain01}
A~Kübler, B.~Kotchoubey, J.~Kaiser, J.~R. Wolpaw, and N.~Birbaumer.
\newblock Brain-computer communication: unlocking the locked in.
\newblock {\em Psychological Bulletin}, 127(3):358--375, 2001.

\bibitem{Keng12}
A.~K. Keng, C.~Z. Yang, C.~Wang, C.~Guan, and H.~Zhang.
\newblock Filter bank common spatial pattern algorithm on bci competition iv
  datasets 2a and 2b.
\newblock {\em Frontiers in Neuroscience}, 6:39, 2012.

\bibitem{2012Filter}
A.~K. Keng, C.~Z. Yang, C.~Wang, C.~Guan, and H.~Zhang.
\newblock Filter bank common spatial pattern algorithm on bci competition iv
  datasets 2a and 2b.
\newblock {\em Frontiers in Neuroscience}, 6:39, 2012.

\bibitem{Lawhern16}
V.~J. Lawhern, A.~J. Solon, N.~R. Waytowich, S.~M. Gordon, C.~P. Hung, and
  B.~J. Lance.
\newblock Eegnet: A compact convolutional network for eeg-based brain-computer
  interfaces.
\newblock {\em Journal of Neural Engineering}, 15(5):056013.1--056013.17, 2016.

\bibitem{Li18}
C.~Li, Z.~Cui, W.~Zheng, C.~Xu, and J.~Yang.
\newblock Spatio-temporal graph convolution for skeleton based action
  recognition.
\newblock 2018.

\bibitem{2002Classification}
A.~Liaw and M.~Wiener.
\newblock Classification and regression by randomforest.
\newblock {\em R News}, 23(23), 2002.

\bibitem{Lotte2018}
Lotte, F., Bougrain, L., Cichocki, A., Clerc, M., Congedo, and Rakotomamonjy.
\newblock A review of classification algorithms for eeg-based brain-computer
  interfaces: a 10 year update.
\newblock {\em Journal of neural engineering}, 2018.

\bibitem{Lu16}
N.~Lu, T.~Li, X.~Ren, and H.~Miao.
\newblock A deep learning scheme for motor imagery classification based on
  restricted boltzmann machines.
\newblock {\em IEEE Transactions on Neural Systems and Rehabilitation
  Engineering}, 2016.

\bibitem{1997Multimodality}
F~Maes and A.~Collignon.
\newblock Multimodality image registration by maximization of mutual
  information.
\newblock {\em IEEE Trans Med Imaging}, 16(2):187--198, 1997.

\bibitem{Makeig12}
S.~Makeig, C.~Kothe, T.~Mullen, N~Bigdely-Shamlo, Z.~Zhang, and
  K~Kreutz-Delgado.
\newblock Evolving signal processing for brain–computer interfaces.
\newblock {\em Proceedings of the IEEE}, 100(13):1567--1584, 2012.

\bibitem{Mcfarland06}
D.~J Mcfarland, C.~W. Anderson, K.~R Muller, A~Schlogl, and D.~J. Krusienski.
\newblock Bci meeting 2005-workshop on bci signal processing: feature
  extraction and translation.
\newblock {\em IEEE Trans Neural Syst Rehabil Eng}, 14(2):135--138, 2006.

\bibitem{J99Designing}
J~Müller-Gerking, G.~Pfurtscheller, and H.~Flyvbjerg.
\newblock Designing optimal spatial filters for single-trial eeg classification
  in a movement task.
\newblock {\em Clinical Neurophysiology}, 110(5):787--798, 1999.

\bibitem{Schirrmeister17}
R.~T. Schirrmeister, J.~T. Springenberg, Ldj Fiederer, M.~Glasstetter,
  K.~Eggensperger, M.~Tangermann, F.~Hutter, W.~Burgard, and T.~Ball.
\newblock Deep learning with convolutional neural networks for eeg decoding and
  visualization.
\newblock {\em Human Brain Mapping}, 2017.

\bibitem{Song19}
Y.~Song, D~Wang, K.~Yue, N.~Zheng, and Zjm Shen.
\newblock Eeg-based motor imagery classification with deep multi-task learning.
\newblock {\em IEEE}, 2019.

\bibitem{Velikovi17}
Petar Velikovi, G.~Cucurull, A.~Casanova, A.~Romero, P~Liò, and Y.~Bengio.
\newblock Graph attention networks.
\newblock 2017.

\bibitem{Woehrle15}
H.~Woehrle, M.~M. Krell, S.~Straube, K.~K. Su, and F.~Kirchner.
\newblock An adaptive spatial filter for user-independent single trial
  detection of event-related potentials.
\newblock {\em IEEE transactions on bio-medical engineering}, 62(7):1696--1705,
  2015.

\bibitem{Wolpaw02}
J.~R. Wolpaw, N.~Birbaumer, D.~J. Mcfarland, G.~Pfurtscheller, and T.~M.
  Vaughan.
\newblock Brain-computer interfaces for communication and control.
\newblock {\em Supplements to Clinical Neurophysiology}, 113(6):767--791, 2002.

\bibitem{Wolpaw91}
J.~R. Wolpaw, D.~J. Mcfarland, G.~W. Neat, and C.~A. Forneris.
\newblock An eeg-based brain-computer interface for cursor control.
\newblock {\em Electroencephalography and Clinical Neurophysiology},
  78(3):252--259, 1991.

\bibitem{2016Identifying}
W.~L. Zheng, J.~Y. Zhu, and B.~L. Lu.
\newblock Identifying stable patterns over time for emotion recognition from
  eeg.
\newblock 2016.

\bibitem{Zhou20}
K.~Zhou, Q.~Song, X.~Huang, D.~Zha, and X.~Hu.
\newblock Multi-channel graph neural networks.
\newblock In {\em Twenty-Ninth International Joint Conference on Artificial
  Intelligence and Seventeenth Pacific Rim International Conference on
  Artificial Intelligence {IJCAI-PRICAI-20}}, 2020.

\end{thebibliography}
\end{document}